\documentclass[runningheads]{llncs}
\usepackage{graphicx}
\usepackage[noend]{algpseudocode}
\usepackage{algorithmicx,algorithm}
\usepackage{amsthm,amsmath,amssymb}
\usepackage{mathrsfs}

\usepackage{color}
\usepackage{multirow}
\usepackage{makecell}
\usepackage{booktabs}
\usepackage{array}
\usepackage{subfigure} 
\usepackage[misc]{ifsym} 
\usepackage[colorlinks,linkcolor=blue]{hyperref}
\begin{document}
	%
	\title{Style Curriculum Learning for Robust \\Medical Image Segmentation}
	\titlerunning{Style Curriculum Learning}
	\author{Zhendong Liu$^{1,2,3}$\thanks{Zhendong Liu and Van Manh contribute equally to this work.}, Van Manh$^{1,2,3\ast}$, Xin Yang$^{1,2,3}$, Xiaoqiong Huang$^{1,2,3}$, Karim Lekadir$^{4}$, Víctor Campello$^{4}$, Nishant Ravikumar$^{5,6}$, Alejandro F Frangi$^{1,5,6,7}$, Dong Ni$^{1,2,3(\textrm{\Letter})}$}
	
	
	\authorrunning{Liu et al.}
	\institute{
		\textsuperscript{$1$}National-Regional Key Technology Engineering Laboratory for Medical Ultrasound, School of Biomedical Engineering, Health Science Center, Shenzhen University, China\\
		\email{nidong@szu.edu.cn} \\
		\textsuperscript{$2$}Medical Ultrasound Image Computing (MUSIC) Lab, Shenzhen University, China\\ 
		\textsuperscript{$3$}Marshall Laboratory of Biomedical Engineering, Shenzhen University, China\\ 
		\textsuperscript{$4$}Artificial Intelligence in Medicine Lab (BCN-AIM), Dept. de Matemàtiques i Informàtica, Universitat de Barcelona, Spain\\
		\textsuperscript{$5$}Centre for Computational Imaging and Simulation Technologies in Biomedicine (CISTIB), School of Computing and School of Medicine, University of Leeds, UK\\ 
		\textsuperscript{$6$}Leeds Institute of Cardiovascular and Metabolic Medicine, University of Leeds, UK\\
		\textsuperscript{$7$}Medical Imaging Research Center (MIRC), KU Leuven, Leuven, Belgium\\
	}
	\maketitle
	
	\begin{abstract}
		The performance of deep segmentation models often degrades due to distribution shifts in image intensities between the training and test data sets. This is particularly pronounced in multi-centre studies involving data acquired using multi-vendor scanners, with variations in acquisition protocols. It is challenging to address this degradation because the shift is often not known \textit{a priori} and hence difficult to model. We propose a novel framework to ensure robust segmentation in the presence of such distribution shifts. Our contribution is three-fold. First, inspired by the spirit of curriculum learning, we design a novel style curriculum to train the segmentation models using an easy-to-hard mode. A style transfer model with style fusion is employed to generate the curriculum samples. Gradually focusing on complex and adversarial style samples can significantly boost the robustness of the models. Second, instead of subjectively defining the curriculum complexity, we adopt an automated gradient manipulation method to control the hard and adversarial sample generation process. Third, we propose the Local Gradient Sign strategy to aggregate the gradient locally and stabilise training during gradient manipulation. The proposed framework can generalise to unknown distribution without using any target data. Extensive experiments on the public M\&Ms Challenge dataset demonstrate that our proposed framework can generalise deep models well to unknown distributions and achieve significant improvements in segmentation accuracy.
		
	\keywords{Image segmentation \and Style transfer \and Curriculum learning}
	\end{abstract}

	\section{Introduction}
	
\begin{figure}
		\centering
		\includegraphics[width=1\textwidth]{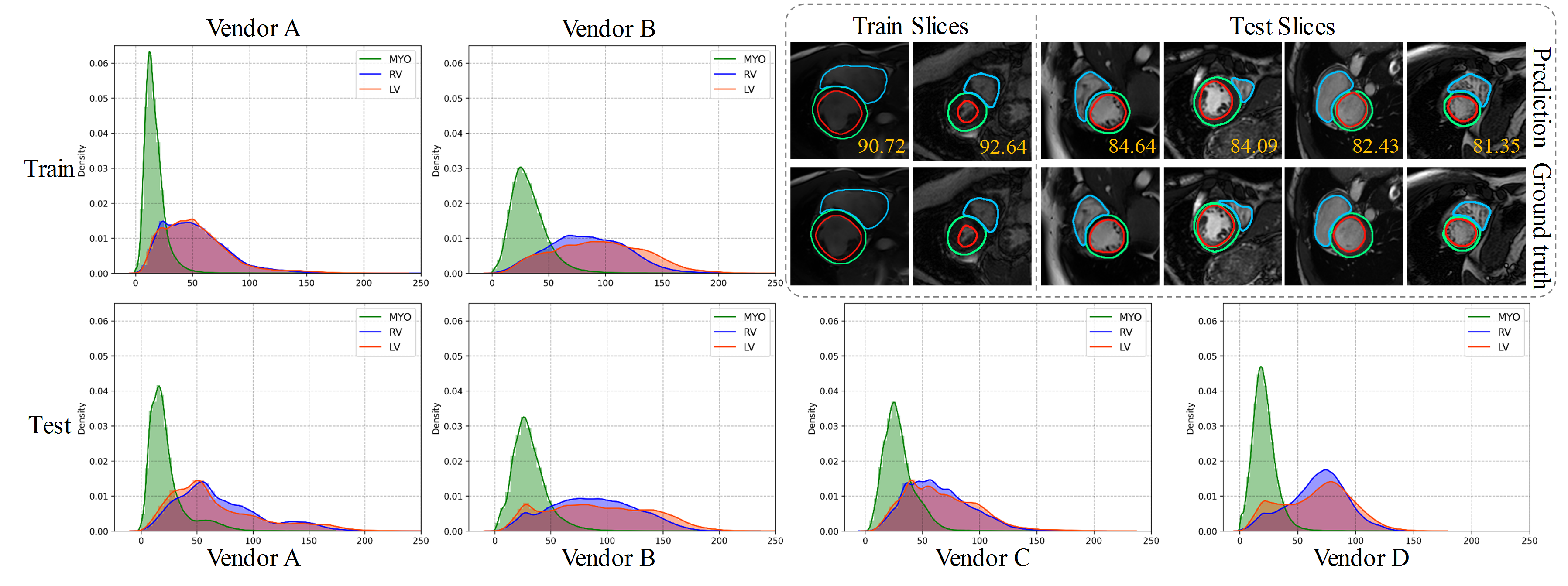}
		\caption{Remarkable distribution shift between the training set and test set across different vendors. The results of the performance drop are shown in the upper right corner. The yellow digits denote the average DSC of the  MYO (green), LV (red)  and RV (blue). } 
		\label{distribution}
	\end{figure}

	Recent studies have witnessed the great success of deep models in medical image segmentation \cite{liu2019deep}. However, these deep models often suffer from a drop in performance when applied to new data distributions, different from the training data (see Fig. \ref{distribution}). It is expensive and practically impossible to collect a large amount of manually annotated data from each new distribution to retrain the model in clinical practice. Therefore, a general and retraining-free framework that ensures model robustness to distribution shifts is highly desired in the clinic. To date, several approaches have been proposed to address this issue.\\ 
	
	\noindent\textbf{Data Augmentation.} Data augmentation (DA) using spatial/intensity transformations is commonly employed to expand the distribution of training data and improve model generalisation. Zhang et al. \cite{zhang2017mixup} proposed \textit{Mixup}, where, two inputs and their corresponding labels are proportionally interpolated to augment the training data. Similarily, Yun et al. \cite{yun2019cutmix} cut and paste local patches from one image into another. Although DA methods can mitigate model overfitting to some degree, they cannot guarantee the ability of deep models to generalise to multi-centre, multi-vendor data, typically encountered in real clinical scenarios.
	
	\noindent\textbf{Domain Adaptation.} GAN-based domain adaptation methods often focus on learning domain-invariant representations \cite{chen2019unsupervised,huang2018multimodal} or aligning the feature space of different distributions \cite{yan2019domain,zhang2018translating} and have shown promise for dealing with variations in data distributions. Curriculum-based domain adaptation methods typically define curriculum complexity subjectively, such as using the average distance of the domain feature space \cite{liu2020open} and developing a ’simpler‘ task than semantic segmentation \cite{zhang2017curriculum}. As these methods rely on the availability of unlabelled target data, they are restricted to the task of domain-mapping/adaptation. Hence, they may not generalise well to new `unseen' data distributions. 
	
	\noindent\textbf{Distribution Generalisation.} Compared to the domain adaptation setting, distribution generalisation methods tested on unseen
	data distributions, are better suited to address the challenges encountered in real clinical scenarios. Some studies employ style transfer methods to remove the distribution shift in test data \cite{liu2020remove,ma2019neural}. Although these approaches are novel, the selection of style data is subjective. Other studies have utilised adversarial training to augment the training set \cite{cai2018curriculum,volpi2018generalizing}. Forcing networks to learn from the most difficult samples can enhance model robustness to a certain degree. Still, it also makes the training more difficult to optimise, even causing catastrophic overfitting \cite{li2020towards}. \par
	
	This work proposes a retraining-free and general framework to improve the ability of segmentation models to generalise to unknown distributions. Our contribution is three-fold. \textit{First}, inspired by the advantages of curriculum learning in improving model generalisation, we design a novel style curriculum, structured in an easy-to-hard adversarial learning strategy, to train the model. Gradually forcing the model to learn from progressively harder adversarial samples allows the model to generalise well to new distributions. Specifically, a style transfer model with style fusion is adopted to generate samples for the curriculum. \textit{Second}, instead of defining the complexity order subjectively, we employ a novel gradient manipulation method to automatically control the sample generation process following the increasingly harder direction. \textit{Third}, since the gradient manipulation-based training is difficult to optimise, we propose Local Gradient Sign (LGS), which locally aggregates the gradient to increase the complexity of the generated samples gradually, thus making the training more stable. Extensive experiments on public data from the Multi-Centre, Multi-Vendor and Multi-Disease Cardiac Segmentation (M\&Ms) Challenge \cite{campello2021multi} show that our method outperforms all 2D methods, and is on par with the top performing approaches of the challenge.
	\begin{figure}[ht!]
		\centering
		\includegraphics[width=0.95\textwidth]{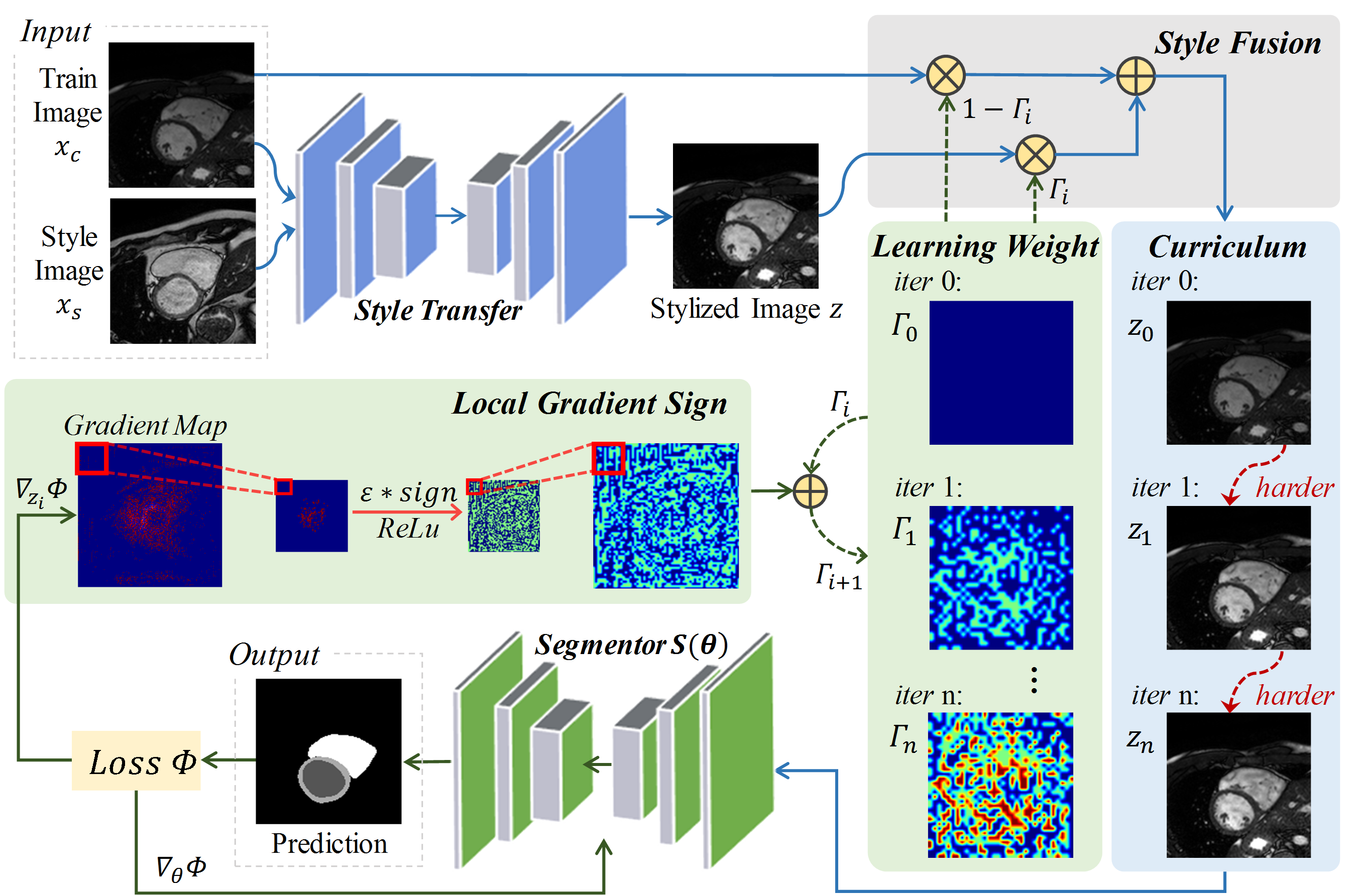}
		\caption{Schematic view of our proposed framework.} \label{framework}
	\end{figure}
	
	\section{Methodology}
	Our framework is shown in Fig. \ref{framework}. First, we adopt a style transfer model with a style fusion operation to generate the curriculum samples $z_i$. Then, we employ the gradient manipulation method to update the learning weight $\Gamma$, so the curriculum samples are arranged from easy to hard. To stabilise the training, we further propose Local Gradient Sign to operate the gradient locally.  \par

	\subsection{Curriculum Learning for Robustness}
	Deep neural networks are typically trained using a sequence of unordered samples. Curriculum learning methods \cite{bengio2009curriculum,hacohen2019power,zhang2017curriculum} can guide the model to learn better by organising the training samples in a meaningful order. A common curriculum needs to address three challenges: (1) Decide the curriculum samples. (2) Arrange the samples in an easy-to-hard order. (3) Ensure model stability during training. Using curriculum learning, deep models can leverage information learned from easy examples, to ease learning of new and harder samples. Gradually concentrating on learning the harder tasks can make the deep model more robust. Creating an effective curriculum is critical to design a reasonable samples generator, a meaningful learning strategy, and a stable learning method. \par

	\subsection{Style Transfer based Sample Generation}
	
	Many works \cite{huang2020style,li2017demystifying} argued that the images' style could be described as the distribution of colours, edges, smoothness, etc. Since style transfer (ST) methods are powerful tools to render one image's style onto another, we employ this approach to produce large sets of stylised images. This expands the training set and increases the richness of the information available for training the segmentation model. ST is only an optional module of our system to generate initial samples.\par
	
	A reasonable samples generator of the curriculum requires the ST method to balance transfer quality and efficiency. Therefore, we follow WaveCT-AIN \cite{liu2020remove} and re-implement their high-quality ST. To better control the stylised degree, we propose a style fusion operation to modulate the content of training images and the style of stylised images. The style fusion weight is updated during back-propagation to control the degree of stylisation. \par
	
	\subsection{Gradient Manipulation based Learning Strategy}
	
	
	\begin{figure}
		\centering
		\includegraphics[width=0.90\textwidth]{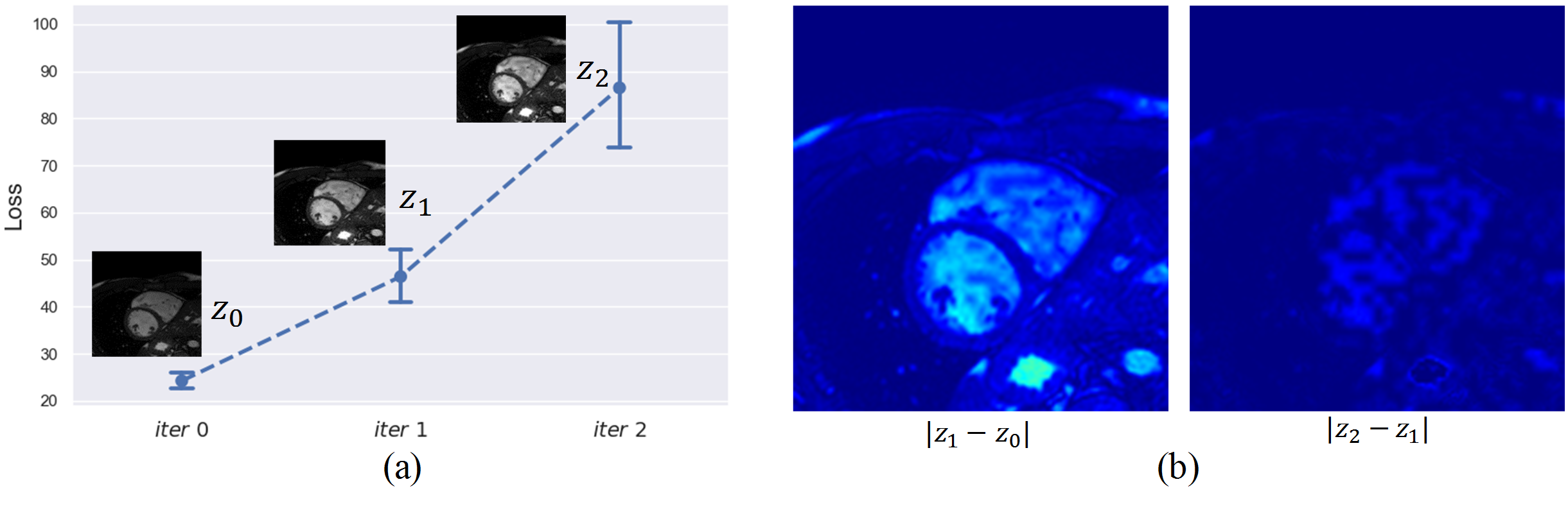}
		\caption{(a) The loss statistics of curriculum samples at different iterations in training. (b) The absolute difference of samples $z_i$ between iter 0 and 1, iter 1 and  2.} 
		\label{harder}
	\end{figure}
	
	ST is a powerful technique that can expand style curriculum samples, but the random transformation lacks control for generating samples in a progressive, easy-to-hard fashion. Following the work from Goodfellow et al. \cite{goodfellow2014explaining},  we adopt a novel gradient manipulation adversarial strategy to phase curriculum learning in increasing order of complexity. They proposed the Fast Gradient Sign Method to generate adversarial examples through the following formula.
	
	\begin{equation}
		x' = x+ \epsilon sign(\nabla_{x} J(f_{\theta}(x),y))
		\label{fgsm}
	\end{equation}
	Where $\theta$ are the parameters of the model, $x$ is the input of the model, $y$ is the label of $x$, $\epsilon$ is a small perturbation vector, $x'$ is the generated adversarial samples, and $J(f_{\theta}(x),y)$ is the loss function used to train the model. The method adjusts the input data by a small step ($\epsilon$) in the gradient direction of the cost function that maximises the loss. With this adversarial approach, the method can quickly generate more difficult samples for the deep models. Similarly, we iteratively accumulate the gradient of the cost function to produce increasingly harder adversarial samples (see step 7 in \textbf{Algorithm 1}). Moreover, in contrast with the typical gradient manipulation adversarial methods \cite{goodfellow2014explaining,li2020towards,volpi2018generalizing}, which add the perturbation into the samples directly, we use it as the learning weight to modulate the training image and stylised image. This ensures control over the adversarial samples generated using our approach (see step 5 in \textbf{Algorithm 1}). To verify the effectiveness of our method, we visualise the loss of three curriculum samples iteratively generated by each training sample (Fig.\ref{harder} (a)) and the image differences of the adjacent curriculum samples (we perform pseudo-color processing in Fig.\ref{harder} (b) for better visualisation). The increase of the training loss shows that the method generates harder samples as the iterations progress.

	\subsection{Local Gradient Smoothing for Stability}
	
	Although the proposed gradient manipulation method guarantees the curriculum's learning order, it is not easy to optimise the model by introducing the whole adversarial perturbations. Therefore, we further propose Local-Gradient-Sign (LGS), to reduce the influence of adversarial perturbations on the model. The LGS formula is defined as follows.
	
	\begin{equation}
		LGS(\Phi, \epsilon, size) = US(ReLu(\epsilon * sign(AP(\nabla_{z_{i}}\Phi, size)))) , 	\quad i=0,1,\cdots, n
		\label{LGS}	
	\end{equation}
	
	Here, $US(\cdot)$ is an UpSample operation, $AP(\cdot)$ is an AvgPooling operation, $z_i$ is curriculum sample and $\epsilon$ is a decimal from 0 to 1. In our curriculum setting, we set the max learning iter $n$ to 3, learning step $\epsilon$ to 0.25 and the pooling size to 4 $\times$ 4. The LGS operation first uses the AvgPooling operation to average the gradient matrix locally for smoothing the perturbations and then performs partial truncation using the ReLu function. Finally, it uses the UpSample operation to restore the original size of the gradient matrix. The proposed LGS operation can alleviate the perturbations to make the training more stable. In our design, the ST and LGS operations are only required during training for data augmentation and attack mitigation of hard samples, respectively. During testing, the trained model is lightweight and efficient without using these two modules. 
	
	\begin{algorithm}[!h]
		\caption{Local Style Curriculum Learning (LSCL)} 
		\hspace*{0.01in} {\bf Notation:}  Segmentation model \textit{$f(\cdot)$}; Style transfer model \textit{$S(\cdot)$}; max  iter \textit{$n$}; learning rate \textit{$\alpha$}; segmentation model parameters $\theta$;  full zeros matrix $O$ and full ones matrix $I$\\
		\hspace*{0.01in} {\bf Input:} 
		Content data \textit{$X_c \in \mathbb{R}^{b \times h \times w} $} and label \textit{$Y_c \in \mathbb{R}^{b \times h \times w} $}; style data \textit{$X_s \in \mathbb{R}^{b \times h \times w}$} \\
		\hspace*{0.01in} {\bf Output:}
		Optimal \textit{$\theta^{*}$}
		\begin{algorithmic}[1]
			\For{$(x_c,y_c)\in (X_c,Y_c)$, $x_s\in X_s$} 
			\State Initialization: $z= S(x_c, x_s)$,  $\Gamma_0 = O\in \mathbb{R}^{h \times w}$
			\For{$i = 0,\ldots, n$} 
			\State Update stylised learning sample: $z_{i}=\Gamma_i * z+(I-\Gamma_i) * x_c$
			\State Compute loss function: $\Phi = J(f(z_{i} , \theta), y_c)$
			\State Accumulate learning weight: $\Gamma_{i+1} = \Gamma_{i} + LGS(\nabla_{z_{i}}\Phi) $
			\State Update model parameters: $\theta \gets \theta-\alpha \nabla_{\theta}\Phi $
			\EndFor
			\EndFor
			\State \Return $\theta$ as $\theta^{*}$
		\end{algorithmic}
	\end{algorithm}
	
	\section{Experimental Results}
	
	Our experimental data came from the M\&Ms Challenge. This challenge cohort had 375 patients scanned in clinical centres using four different magnetic resonance scanner vendors (A, B, C and D). The training set contained 150 images from two different vendors (75 each of vendor A and B), in which only the end-diastolic (ED) and end-systolic (ES) phases are annotated.  The images have been segmented by experienced clinicians from the respective institutions, including contours for the left (LV), right ventricle (RV) blood pools, and the left ventricular myocardium (MYO). All experiments are evaluated on 50 new studies from vendor A, B, C and 50 additional studies from the unseen vendor D. \par 
	
	We totally obtained 3,284 training slices from the annotated ED \& ES phases and 10,607 style slices cross the short-axis view of vendor A and B. We first used the training slices to train the segmentation model \cite{ronneberger2015u}, employing Adam optimiser (learning rate of $10^{-3}\sim10^{-5}$). Based on the well-trained segmentation model and pre-trained style transfer model, we randomly sampled the slices of vendors A, B as content-style input. Then we adopted the LSCL algorithm to finetune the segmentation model for 10 epochs using SGD-momentum optimiser (learning rate of $10^{-5}$). To further improve the model robustness, we finally adopted the Test-time Augmentation (TTA), including three rotation operations ($90^{\circ}$, $180^{\circ}$, $270^{\circ}$) to aggregate the multiple predictions. We used a segmentation loss comprising cross-entropy loss (weight of 0.6) and dice loss (weight of 0.4). \par
	
	\subsection{Result Details}
	
	We performed the comparative experiment and ablation experiment using indicators of Dice similarity coefficient (DSC), Jaccard index (JAC), Hausdorff distance (HD, [mm]) and Average symmetric surface distance (ASSD, [mm]). In all experiments, we adopted the same evaluation criteria and ranking method as the M\&Ms Challenge performs \cite{campello2021multi}, including the \textit{Min-max Score}. Due to the space limitations, we computed the averaged metrics over the LV, RV and MYO. \par
	
	\begin{table*}[!htbp]
		\centering
		\begin{center}
			\caption{The mean (std) results of DSC and HD on different vendors of all patients.}
			\scalebox{0.74}{\setlength{\tabcolsep}{1mm}{
					\begin{tabular}{l|cc|cc|cc|cc|ccc}
						\toprule[1pt]
						\multirow{2}{*}{\textbf{Methods}} & \multicolumn{2}{c|}{Vendor A} &
						\multicolumn{2}{c|}{Vendor B} & \multicolumn{2}{c|}{Vendor C} & \multicolumn{2}{c|}{Vendor D} &
						\multirow{2}{*}{\shortstack{\textbf{DSC}\\\textbf{Score}}}
						&
						\multirow{2}{*}{\shortstack{\textbf{HD}\\\textbf{Score}}} 
						
						& \multirow{2}{*}{\shortstack{\textbf{Min-max}\\\textbf{Score}}} 
						\\
						\cline{2-9}
						
						\specialrule{0em}{0.8pt}{0.5pt}
						& DSC $\uparrow$ & HD $\downarrow$  & DSC $\uparrow$ & HD $\downarrow$  & DSC $\uparrow$ & HD $\downarrow$ & DSC $\uparrow$ & HD $\downarrow$   \\
						\cline{1-12}
						\multirow{2}{*}{U-net} & 0.858 & 17.113  & 0.835 & 13.392 & 0.822 & 17.616 & 0.835 & 17.153 & \multirow{2}{*}{0.8345}& \multirow{2}{*}{16.674}& \multirow{2}{*}{0.000} \\
						& (0.034) &  (13.218) & (0.056) & (7.541) & (0.067) & (13.544) & (0.045) & (14.328)  \\
						
						\cline{1-12}
						
						\multirow{2}{*}{Mixup} & 0.872 & 10.441  & 0.859 & 11.483 & 0.837 & 12.596 & 0.837 & 11.904 & \multirow{2}{*}{0.8465}& \multirow{2}{*}{11.821}& \multirow{2}{*}{0.528}  \\
						& (0.034) &  (3.315) & (0.047) & (3.605) & (0.065) & (6.010) & (0.040) & (4.506)   \\
						
						\cline{1-12}
						
						\multirow{2}{*}{P3} & 0.878 & 12.587  & 0.887 & 9.872 & 0.865 &  10.461 & 0.864 & 13.969 & \multirow{2}{*}{0.8705}& \multirow{2}{*}{11.887}& \multirow{2}{*}{0.779}  \\
						& (0.042) &  (12.279) & (0.053) & (3.607) & (0.055) & (6.499) & (0.050) & (16.845)  
						\\

						\cline{1-12}
						
						\multirow{2}{*}{LSCL} & 0.877 & 9.888  & 0.865 & 10.746 & 0.857 & 11.224 & 0.866 & 11.479 & \multirow{2}{*}{0.8647}& \multirow{2}{*}{11.007}& \multirow{2}{*}{0.789}  \\
						& (0.033) &  (3.042) & (0.051) & (3.315) & (0.063) & (5.445) & (0.036) & (5.027) 
						\\
						
						\cline{1-12}
						
						\multirow{2}{*}{P2} & 0.884 & 12.461  & 0.892 & 9.796 &  0.870 & 9.548 & 0.867 & 13.432 & \multirow{2}{*}{0.8750}& \multirow{2}{*}{11.370}& \multirow{2}{*}{0.869}  \\
						& (0.039) &  (12.237) & (0.051) & (3.369) & (0.046) & (3.294) & (0.049) & (14.845)  \\

						\cline{1-12}
						
						\multirow{2}{*}{\textbf{LSCL-TTA}} &  0.884 & 9.802  & 0.876 & 10.238 & 0.865 & 10.592 & 0.868 & 11.195 & \multirow{2}{*}{0.8710} & \multirow{2}{*}{10.602}& \multirow{2}{*}{0.890} \\
						& (0.032) &  (3.529) & (0.047) & (2.979) & (0.059) & (5.192) & (0.034) & (4.890)  \\
						
						\cline{1-12}
						
						\multirow{2}{*}{P1} & 0.889 & 12.072  & 0.893 & 9.482 & 0.876 & 9.465 & 0.877 & 13.091 & \multirow{2}{*}{0.8813} & \multirow{2}{*}{11.111}& \multirow{2}{*}{0.958} \\
						& (0.042) &  (12.641) & (0.046) & (3.343) & (0.042) & (3.562) & (0.042) & (14.838)  \\
						
						\toprule[0.8pt]
			\end{tabular}}}
			\label{compare}
		\end{center}
	\end{table*}
	
	\begin{table*}[!htbp]
		\centering
		\begin{center}
			\caption{Training and inference time of different methods.}
			\scalebox{0.83}{\setlength{\tabcolsep}{1mm}{
					\begin{tabular}{l|c|c|c}
						\hline
						
						Method & Training time  & Inference time & GPU Device  \\
						\hline
						
						\textbf{LSCL-TTA} & 5 h & 0.2 s & GTX 1080 Ti \\				
						P1      & 60 h  & $ \approx 1 s $ & Titan XP  \\
						P2      & 48 h  & 4.8 s & Tesla V100  \\	
						P3       & 4-5 days & N/A & TITAN V100  \\

						\toprule[0.8pt]
			\end{tabular}}}
			\label{time}
		\end{center}
		
	\end{table*}

	\begin{table}[!h]
	\centering
	\begin{center}
		\caption{Ablation experiment on ED and ES volumes.}
		\scalebox{0.81}{\setlength{\tabcolsep}{1mm}{
				\begin{tabular}{l|cccc|cccc}
					\hline
					\multirow{2}{*}{\textbf{Methods}}
					& \multicolumn{4}{c|}{ED} & \multicolumn{4}{c}{ES} \\
					\cline{2-9}
					\specialrule{0em}{1pt}{0.1pt}
					& DSC $\uparrow$ & JAC $\uparrow$ & HD $\downarrow$ & ASSD $\downarrow$ & DSC $\uparrow$ & JAC $\uparrow$ & HD $\downarrow$ & ASSD $\downarrow$ \\
					\hline
					WaveCT-AIN  & 0.847 & 0.747 & 11.469 & 1.146 & 0.827 & 0.713 & 12.386 & 1.413 \\
					
					SCL      & 0.861 & 0.766 & 10.743 & 1.038 & 0.843 & 0.736 & 11.042 & 1.159 \\
					
					LSCL     & 0.876 & 0.787 & 10.543 & 0.920 & 0.853 & 0.751 & 11.459 & 1.146 \\
					
					\textbf{LSCL-TTA} & {\color{blue} 0.882} & {\color{blue} 0.797} & {\color{blue} 10.246} & {\color{blue} 0.870} & {\color{blue} 0.861} & {\color{blue} 0.762} &  {\color{blue} 10.898} & {\color{blue} 1.077}  \\
					\toprule[0.8pt]
		\end{tabular}}}
		\label{ablation}
	\end{center}
\end{table}
	\noindent \textbf{Method Comparison.}
	The quantitative comparisons among our methods and others on M\&Ms data are shown in Tab.\ref{compare}. U-net is the baseline and Mixup is a common data augmentation method. P3-P1 are the top three methods on the leaderboard of the M\&Ms Challenge. Tab.\ref{compare} shows our method helps the U-net significantly improve the performance by 3.65\% and 6.072 mm in terms of the DSC and HD metrics. Smaller standard deviations show that our method is more robust and stable against distribution shifts. Tab.\ref{compare} and Tab. \ref{time} show that our proposed method (LSCL-TTA) achieves competitive performance among the top ranking methods, while presents the lowest computational costs. \par

	\begin{figure}[!h]
		\centering
		\includegraphics[width=0.91\textwidth]{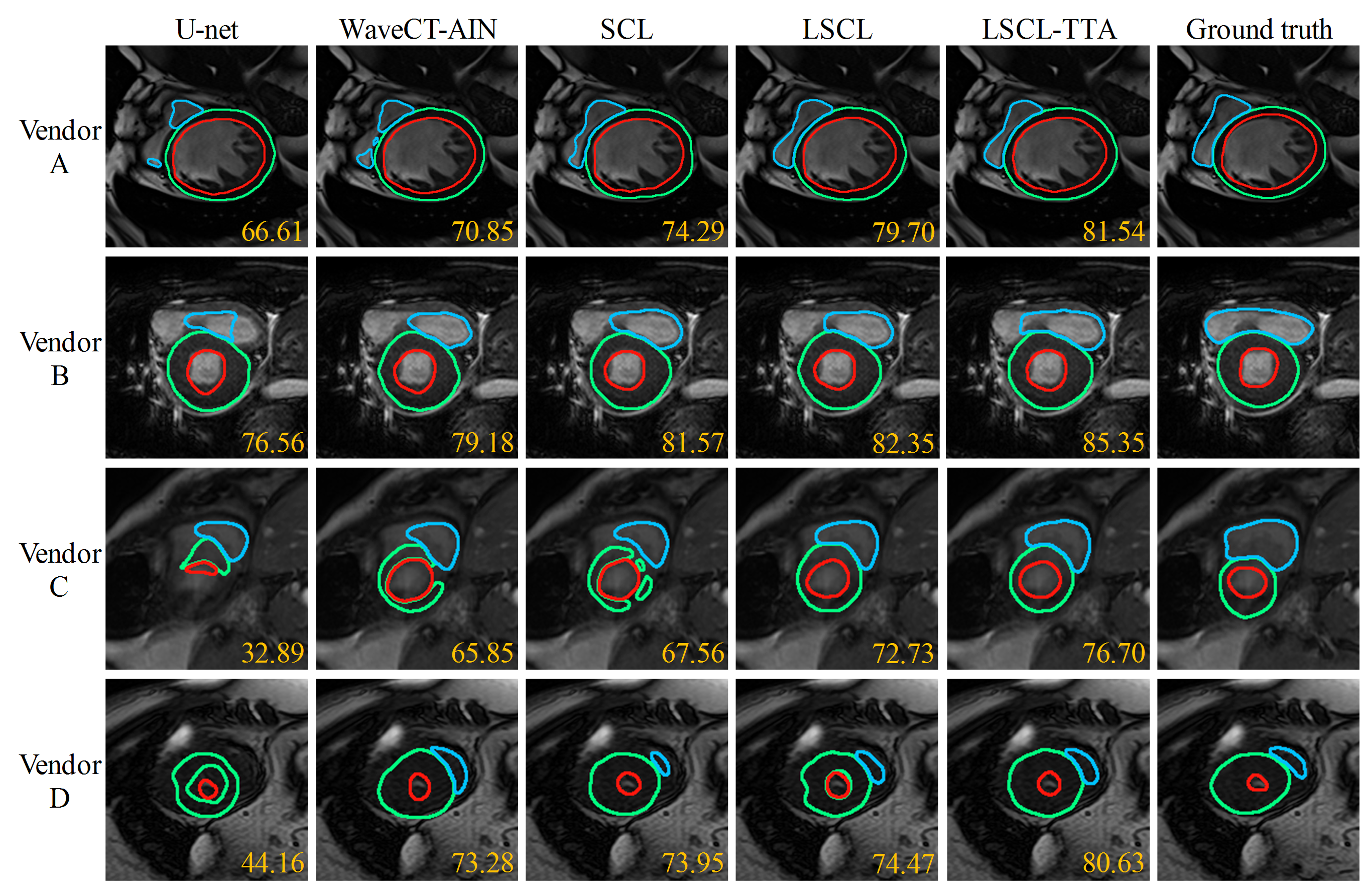}
		\caption{Visualisation of ablation methods at different vendors. } 
		\label{vis}
	\end{figure}
	\noindent \textbf{Ablation Experiment.} 
	To thoroughly evaluate our proposed framework, we conducted two ablation experiments. The first experiment uses only the style transfer method \textit{WaveCT-AIN} \cite{liu2020remove} to generate random style transformations and then feeds the stylised outputs to train the segmentation model. The second experiment does not introduce the LGS method, which we name SCL. As shown in Tab. \ref{ablation}, the WaveCT-AIN shows the worst performance since the model learns the stylised samples arbitrarily, causing the model to be trapped in local minima. SCL, instead, can make the model learn increasingly harder tasks. We further proposed LSCL to stabilise the adversarial training. Fig.\ref{vis} visualises the segmentation results of test samples corresponding to different ablation methods. The proposed LSCL-TTA shows the best segmentation results on each vendor. \par

	\section{Conclusion}
	This paper proposes a novel style curriculum learning framework to ensure segmentation models are robust against the distribution shift. Extensive experiments show that our proposed framework can significantly improve the generalisation. The proposed framework is universal and retraining-free, which makes it compelling for use in clinical practice.
	\\
	
	\noindent \textbf{Acknowledgement.} This work was supported by the National Key R\&D Program of China (No. 2019YFC0118300), Shenzhen Peacock Plan (No. KQTD20160\\-53112051497, KQJSCX20180328095606003), Royal Academy of Engineering under the RAEng Chair in Emerging Technologies (CiET1919/19) scheme, EPSRC TUSCA (EP/V04799X/1), the Royal Society CROSSLINK Exchange Programme (IES/NSFC/201380), European Union’s Horizon 2020 research and innovation program under grant agreement number 825903 (euCanSHare project), Spanish Ministry of Science, Innovation and Universities under grant agreement RTI2018-099898-B-I00. 
	%
	%
	%
	\bibliographystyle{splncs04}
	\bibliography{refs}

\end{document}